\documentclass[a4paper,11pt]{article}
\usepackage{pos}

\title{The Terzina instrument onboard the NUSES space mission}
 \ShortTitle{The Terzina instrument onboard NUSES}

\author*[a,b]{R. Aloisio}
\author[c]{L. Burmistrov}
\author[a,b]{A. Di Giovanni}
\author[c]{M. Heller}
\author[c]{T. Montaruli}
\author[c]{C. Trimarelli}

\affiliation[a]{Gran Sasso Science Institute, viale F. Crispi 7, L'Aquila, Italy}

\affiliation[b]{INFN - Laboratori Nazionali Gran Sasso, via G. Acitelli 22, Assergi (AQ), Italy}

\affiliation[c]{Département de Physique Nuclèaire et Corpusculaire, Université de Genève, 1205 Genève, Switzerland}
\onbehalf{for the NUSES collaboration} 

\emailAdd{roberto.aloisio@gssi.it}
\abstract{
In this paper we will introduce the Terzina instrument, which is one of the two scientific payloads of the NUSES satellite mission. NUSES serves as a technological pathfinder, hosting a suite of innovative instruments designed for the in-orbit detection of cosmic rays, neutrinos, and gamma rays across various energy ranges. The Terzina instrument itself is a compact telescope equipped with Schmidt-Cassegrain optics. Its primary objective is to detect Cherenkov radiation emitted by Extensive Air Showers generated by the interaction of high-energy (> 100 PeV) cosmic rays with the Earth's atmosphere. Terzina represents a critical step forward in the development of future space-based instruments aimed at detecting upward-moving showers induced by tau-leptons and muons resulting from the interaction of high-energy astrophysical neutrinos with the Earth. In this paper, we will delve into the key technical aspects of the Terzina instrument, its capabilities, and its potential for detection.
}

\ConferenceLogo{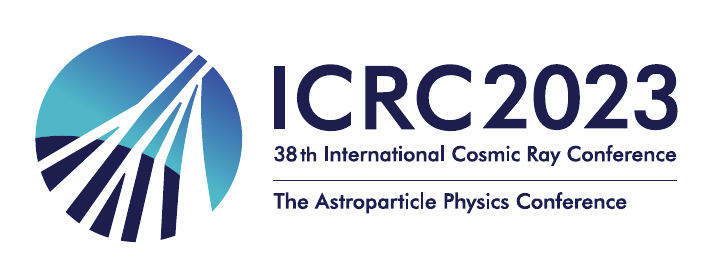}

\FullConference{%
38th International Cosmic Ray Conference (ICRC2023)\\
  26 July - 3 August, 2023\\
  Nagoya, Japan}

\begin{document}
\maketitle

\section{Introduction}
The NUSES (Neutrinos and Seismic Electromagnetic Signals) satellite mission is a collaborative project led by the Gran Sasso Science Institute (GSSI) aimed at exploring new scientific and technological pathways for future astroparticle physics space-based detectors. This project is conducted in collaboration with the Istituto Nazionale di Fisica Nucleare (INFN), the Italian Space Agency (ASI), several Universities in Italy, the University of Geneva in Switzerland and the University of Chicago in the USA. The NUSES mission is supported by Thales Alenia Space Italy (TAS-I), industrial partner providing the satellite platform 2MF/NIMBUS (New Italian Micro BUS) with a modular and flexible design based on additive manufacturing techniques. The NUSES satellite, scheduled to launch in the second half of 2025 under the management of ASI, will be a ballistic mission without orbital control, operating at a Low Earth Orbit (LEO), with an altitude at the Beginning of Life (BoL) of 535 km, with a high inclination of 97.8° (LTAN = 18:00) in a Sun-synchronous orbit along the day-night boundary. The nominal duration of the NUSES mission is three years (End of Life, EoL).

The NUSES satellite will host two innovative scientific payloads: Zirè and Terzina. %Zirè consists of a scintillating fiber tracker, a stack of plastic scintillator counters, an array of LYSO crystals, an active veto system, and a Low Energy Module (LEM). It will perform spectral measurements of electrons, protons, and light nuclei below few hundreds of MeV. Zirè will also test new detection tools for 0.1-10 MeV photons, monitor the Van Allen radiation belts and space weather effects. 
In this proceedings paper, we will focus on the Terzina instrument, while detailed information about the Zirè detector can be found in \cite{ZireProc}. 

Terzina is a telescope specifically designed for the detection of Cherenkov light emitted by Extensive Air Showers (EAS) induced by high-energy Cosmic Rays (CR) and neutrinos in the Earth's atmosphere. In astrophysical environments, high-energy neutrinos are produced through the decay chain of pions, leading to an equipartition (due to flavour oscillation) among the three different leptonic flavours when observed at the Earth. At sufficiently high energy ($E> 1~PeV$), tau neutrinos and, to a lesser extent, mu neutrinos passing through the Earth can produce $\tau$ and $\mu$ leptons, which can emerge by decaying or interacting in the atmosphere when the elevation angles of the neutrino momentum on the Earth's surface are small (Earth skimming events). As a result, Earth skimming neutrinos generate EAS moving in the atmosphere from bottom to top \cite{Austin}, similar to the EAS produced by charged particles (CR) impinging the atmosphere from above the Earth limb \cite{Austin}. The Cherenkov emission from these EAS can be detected by space-based instruments, providing a unique signal for Low Earth Orbit (LEO) satellites \cite{POEMMA,CHANT}, which, given the high exposures, could potentially revolutionise the observation of high-energy neutrinos and CR. Terzina serves as a technological pathfinder, aiming to detect the EAS Cherenkov emission demonstrating the viability of the space-based detection technique. 

This paper presents a general description of the Terzina telescope and its observational capabilities. 
%The organisation of the paper is as follows: in Section \ref{sec:Cherenkov}, we discuss the expected Cherenkov signal on the telescope; Section \ref{sec:Telescope} focuses on the telescope and its Focal Plane Assembly (FPA), while more detailed information about the sensors and front-end electronics can be found in \cite{UHECR22,TerzinaProc}. Finally, in Section \ref{sec:Conclusions}, we conclude by discussing the expected detection potential of the Terzina instrument.

\section{Cherenkov emission observed from space}
\label{sec:Cherenkov}

\begin{figure}[!h]
\centering
\includegraphics[scale=.35]{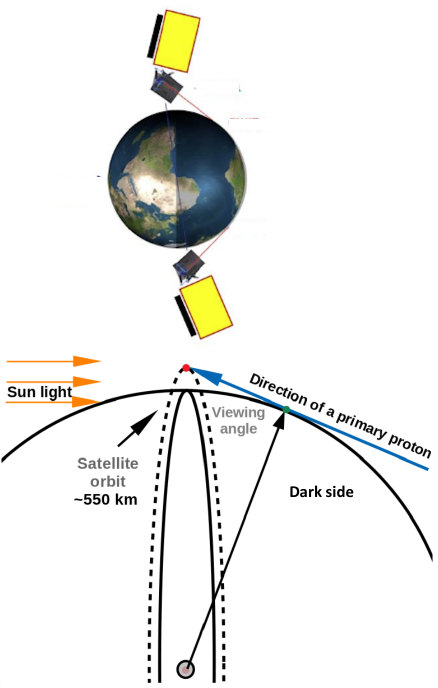}
\includegraphics[scale=.63]{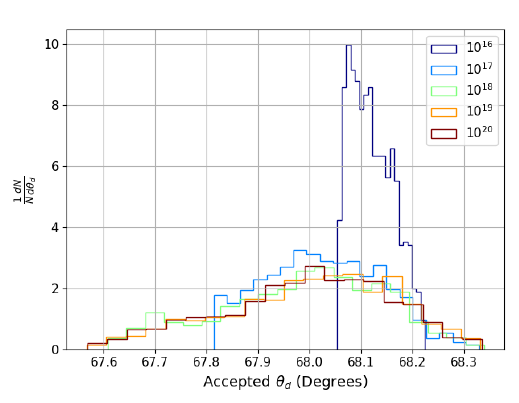}
\includegraphics[scale=.41]{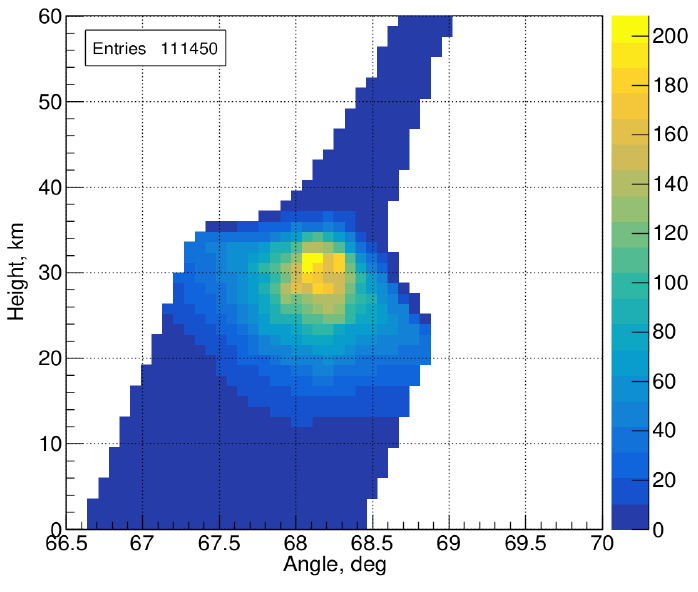}
\caption{[Left panel] Schematic of the orbital configuration and the geometry of an above-the-limb event. [Central panel] Distribution of the relevant line of sight angles for different values of the proton EAS energy (as labelled). [Right panel] Distribution of the Cherenkov photons produced by a proton EAS of 100 PeV energy as function of the viewing angle and above the limb altitudes of the first interaction point.}
\label{fig1}  
\end{figure}

The Cherenkov emission produced by an EAS is mainly due to the high energy ($E>100$ MeV) electron-positron pairs generated in a large amount during the shower development. Thus, the number of Cherenkov photons emitted by an EAS is directly proportional to the shower energy, which corresponds to the energy of the primary particle that initiated the cascade. Considering the specific characteristics of the Terzina telescope, particularly the area of its primary mirror (approximately $0.1~m^2$), as discussed in the next section, it is capable of detecting the Cherenkov emission only from EAS with energies exceeding 100 PeV. Consequently, Terzina is expected to predominantly observe CR with trajectories above the Earth's limb due to their higher flux compared to neutrinos (roughly 4 orders of magnitude higher). Nevertheless, the Cherenkov signal produced by these CR events, apart from the incoming direction, exhibits nearly identical properties to the expected signal from neutrino events occurring below the limb, such as similar wavelength spectra of the arriving photons, as well as comparable spatial profiles and time distributions. Thus, above-the-limb CR events serve as a reliable benchmark for directly testing the various components of an in-orbit Cherenkov telescope (e.g., optics, photo-sensor, electronics, and triggers) during the actual mission. This strategic approach underpins the Terzina mission, which aims to validate the detection technique through in-orbit testing.

In this section we will briefly review the nature of Cherenkov emission as observed from a space based telescope in the case of above-the-limb EAS. The results presented are based on the EASCherSim computational framework ({\url c4341.gitlab.io/easchersim/index.html}), a simulator designed for modelling the production and atmospheric transport of Cherenkov photons by EAS, build upon the findings discussed in \cite{Austin}. To provide an estimate of the expected signal in the Terzina telescope, we consider the case of EAS generated by protons. Due to the geometry of above-the-limb trajectories, a significant portion of the particle cascade occurs at high altitudes in a rarefied atmosphere. Consequently, the generation of optical Cherenkov emission is limited, but so is its atmospheric attenuation during photon propagation. Therefore, a detailed calculation of the Cherenkov signal strength and geometry is necessary to determine the overall instrumental sensitivity to such events.

\begin{figure}[!h]
\centering
\includegraphics[scale=.415]{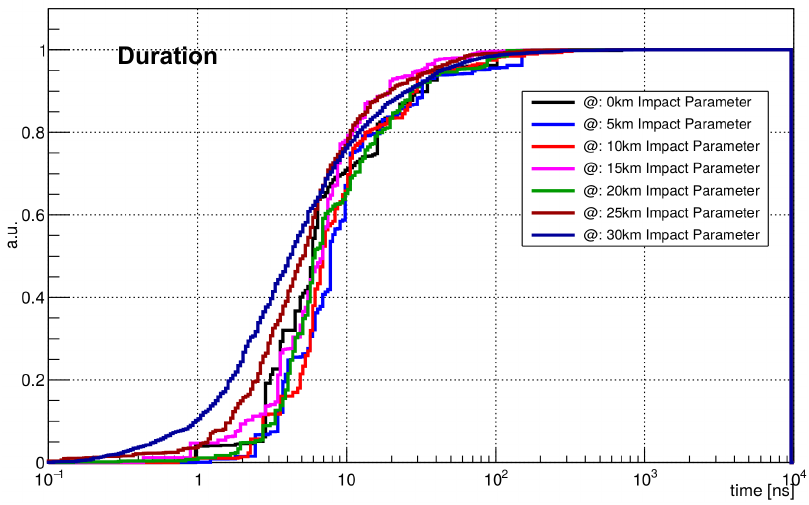}
\includegraphics[scale=.395]{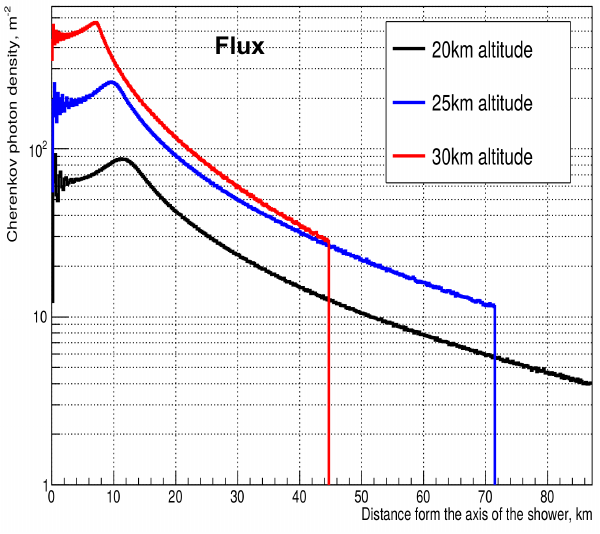}
\includegraphics[scale=.395]{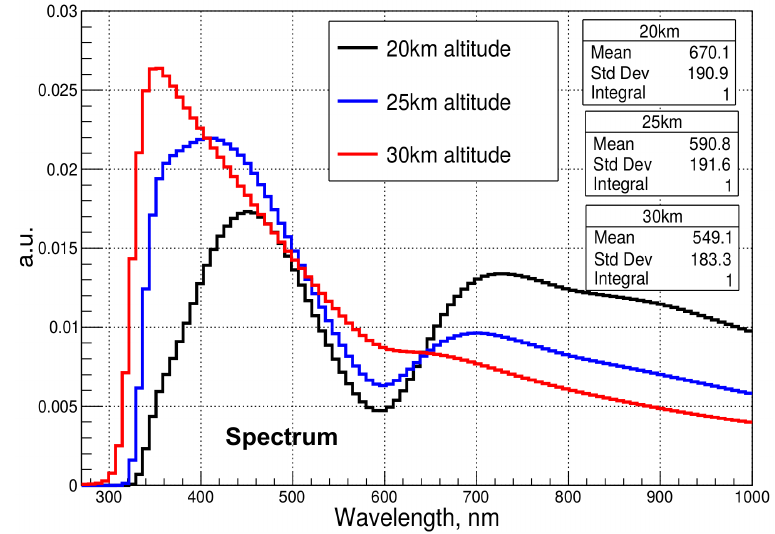}
\caption{[Left panel] Temporal evolution of the Cherenkov burst of a 100 PeV proton EAS for different values of the altitude of the first interaction point (as labelled). [Central panel] Total flux of Cherenkov photons at the Terzina (BoL) altitude integrated over 100 ns as function of the distance from the EAS axis, in the case of a proton EAS of 100 PeV energy for different altitudes of the first interaction point of the proton (as labeled). [Right panel] Spectrum of Cherenkov photons produced by a proton EAS of 100 PeV energy at the Terzina (BoL) altitude for different altitudes of the first interaction point of the proton (as labeled).}
\label{fig2}  
\end{figure}

The results presented in figures \ref{fig1}, \ref{fig2} show several important points. Given the geometry of the observation from the Terzina altitude (figure \ref{fig1} left panel) and the characteristics of the atmosphere, Cherenkov emission can be observed from a tiny layer of the atmosphere, with a angular size less than 1$^\circ$, which corresponds to altitudes above the Earth's limb that span from 20 km up to 50 km (the Earth's limb is seen by Terzina at an elevation angle $\theta_d=67.5^\circ$), as follows from the central and right panels of figure \ref{fig1}. The Cherenkov signal is a burst with a typical duration of few tens on nano seconds (left panel figure \ref{fig2}) of visible-UV photons distributed on a cone with a very narrow aperture ($\delta\simeq 1^\circ$) around the EAS axis, which corresponds to the direction of motion of the primary particle that generated the EAS. At the operative altitudes of Terzina the cone has a typical base radius of few tens of km, with a flux integrated over the burst duration of about 100 photons per $m^2$, in the case of a proton EAS with 100 PeV energy (central panel figure \ref{fig2}). Finally, it is interesting the effect of photons propagation across the Earth's atmosphere that mainly suffer the effect of absorption on the ozone layers that reduce the photons spectra between wavelength of 500 nm and 700 nm, with a progressive attenuation of this effect for EAS generated at increasing altitudes (due to the reduction of the ozone layer traversed), as follows from the right panel of figure \ref{fig2}.

\section{The Terzina instrument}
\label{sec:Telescope}

\begin{figure}[!h]
\centering
\includegraphics[scale=.203]{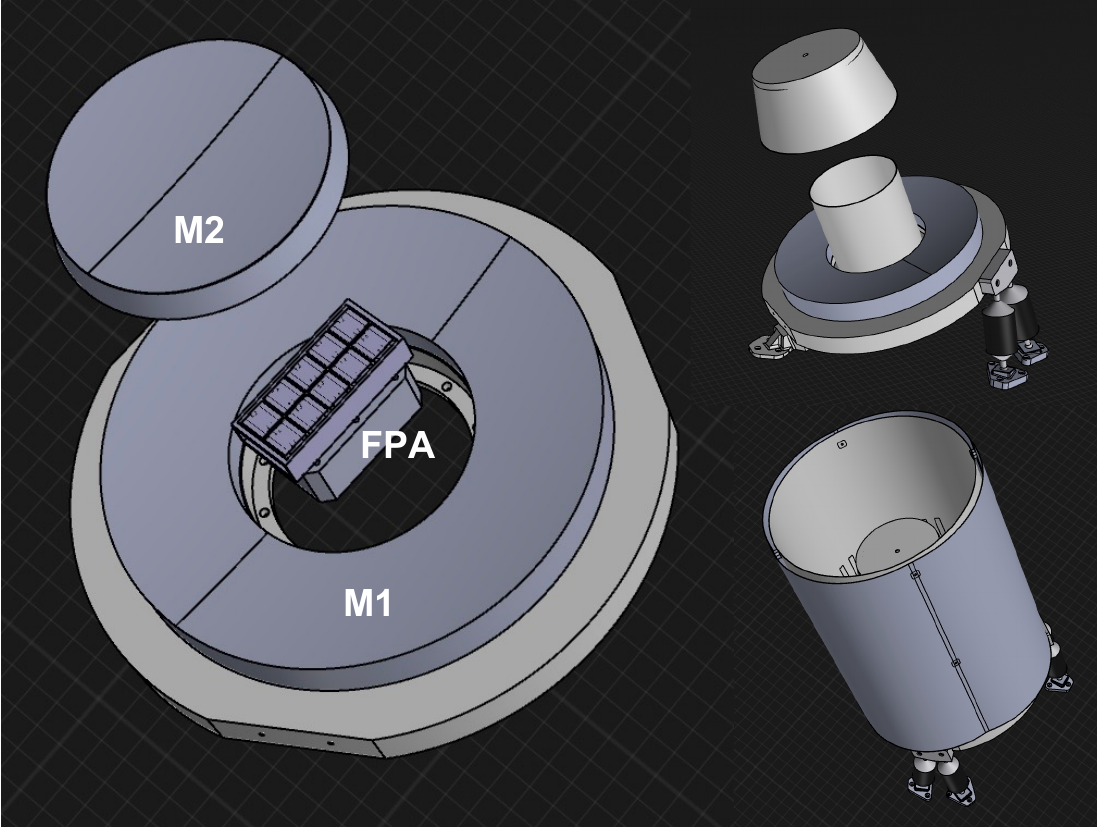}
\includegraphics[scale=.375]{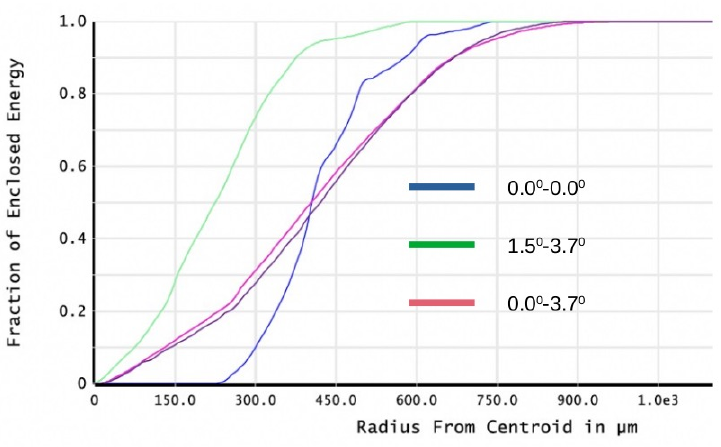}
\includegraphics[scale=.285]{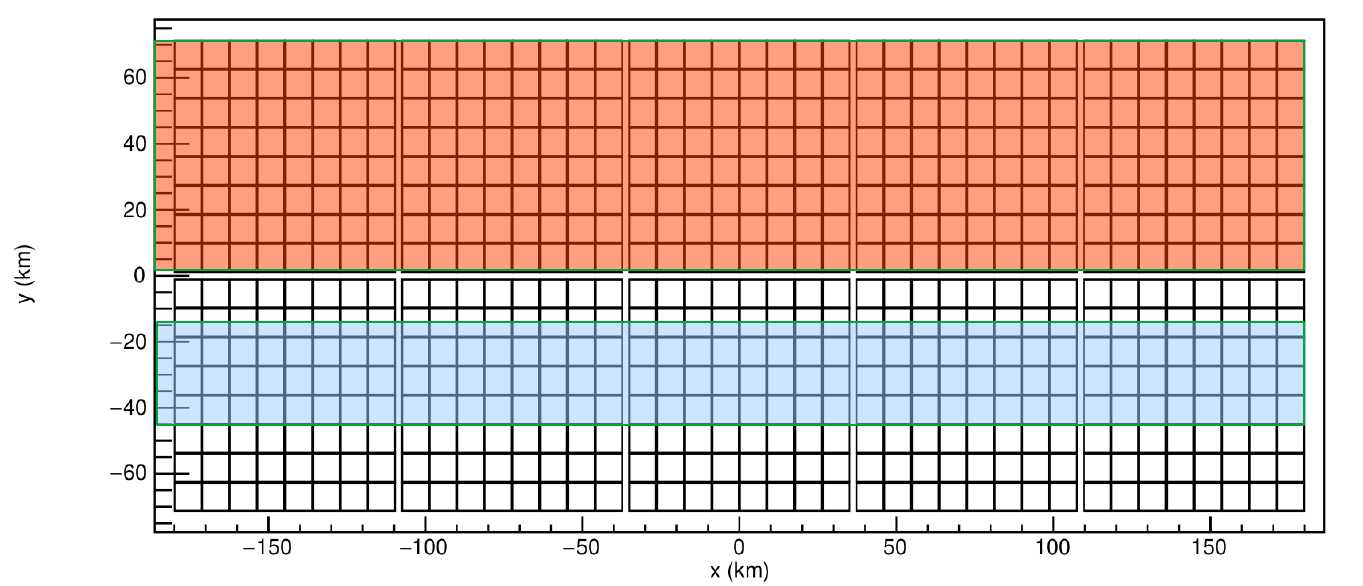}
\caption{[Left panel] Scheme of the optics (M1, M2 primary and secondary mirror) and the FPA of Terzina with the baffles structure to protect from straight light pollution. [Central panel] Point spread function with the encircled energy on the FPA produced by photons for different photons incidence angles respect to the telescope focal axis (as labelled). [Right panel] Scheme of the focal plane with the pixels structure, coloured bands show the corresponding observed regions below (red band) and above (blue band) the limb. The axis label the corresponding length in km probed at the Earth along the limb line (x axis) and across it (y axis).}
\label{fig3}  
\end{figure}

The Terzina detector is composed by a near-UV-optical telescope, with a Schmidt-Cassegrain optics, and the Focal Plane Assembly (FPA), figure \ref{fig3} left panel. The optical system of the telescope is based on a dual mirror configuration composed of two parabolic mirrors: primary, with radius 394~mm, and secondary, with radius 144~mm, placed at a relative distance of 280~mm. The FPA has a maximum radius 121 mm and it is placed at a distance of 40~mm from the primary mirror. This configuration is chosen to maximise the focal length up to 925~mm in a compact telescope which, including the baffles needed to obscure straight light propagation on the FPA (left panel figure \ref{fig3}), should fit in an envelope of 600x600x730 mm$^3$. The telescope will operate inclined by $67.5^\circ$ with respect to nadir, with the optical axis pointing towards the dark side of the Earth’s limb, the expected duty cycle is around $40\%$. The star tracker system of the satellite platform maintains the optical axis configuration with a high accuracy of 0.1$^\circ$. The total weight of the Terzina instrument (telescope and FPA) is around 35~kg.

The FPA is designed to detect photons from both below and above the limb. It has a rectangular shape with a $2 : 5$ aspect ratio. It is composed of 10 Silicon Photon Mutlipliers (SiPM) arrays \cite{SiPM} of $8\times 8$ pixels forming 2 rows of 5 arrays each (640 pixels overall, see figure \ref{fig3} right panel). Given the Schmidt-Cassegrain optics the upper row of 5 SiPM arrays will observe events coming from below the Earth's limb (red area in right panel of figure \ref{fig3}), this part of the FPA will perform a clear characterisation of the background and is unlikely to observe neutrino-induced EAS. On the other hand, the lower row of 5 SiPM arrays will observe events coming from EAS generate by CR from above the limb, with the blue area in the right panel of figure \ref{fig3} signalling the most contributing part of the atmosphere. The axis in the right panel of figure \ref{fig3} show the length probed by the telescope at the Earth, along the limb line and across it. Terzina observes a vast volume of the atmosphere with a section across the Earth's limb of 360 × 140 km$^2$. Given the focal length of the telescope $F_l=925~mm$ and the SiPM pixels size $r_{SiPM}\simeq 3~mm$, the Field-of-View (FoV) per pixel of the FPA can be estimated as ${\rm FoV}_{pix}=\arctan(r_{SiPM}/F_l)\simeq 0.18^\circ$, with a telescope FoV of 7.20$^\circ$ (40 pixels) along the Earth's limb and 2.88$^\circ$ across it (16 pixels). The point spread function (PSF) of the Terzina optical system is compatible with the 3 × 3 mm$^2$ pixel size chosen for the FPA, with the overall encircled energy always contained inside 1.5~mm independently of the inclination angle of the incoming photons (figure \ref{fig3} central panel). 

The SiPM sensors are provided by the Fondazione Bruno Kessler (FBK) and briefly described below (see \cite{UHECR22} for a detailed discussion). The camera frontend electronics is composed of 10 Application Specific Integrated Circuits (ASICs) \cite{TerzinaProc}, each reading one SiPM array with $8\times 8$ channels. The ASIC has an input amplification stage and digitises signals, upon validation of trigger conditions, as determined by the trigger logic implemented in the ASICs and in a dedicated Field Programmable Gate Array (FPGA). The ASIC digitises the signal on a programmable time interval (see below) that spans from a minimum of 180 ns up to 1.280 $\mu$s, enabling pulse-shape reconstruction \cite{TerzinaProc}.

In order to build a complete simulation chain of the Terzina detector it is important to estimate the expected background. This is composed by: the Night Glow Background (NGB) of visible light and the background radiation of charged particles in the 100~keV - 100~MeV energy range. The rate per pixel due to the NGB ($R_{pix}^{NGB}$) has been estimated using the formula \cite{NGB}: $R_{pix}^{NGB} = \eta \times \Delta\Omega \times \phi_{NGB} \times S \times PDE_{eff}$ where: 
\begin{figure}[!h]
\centering
\includegraphics[scale=.58]{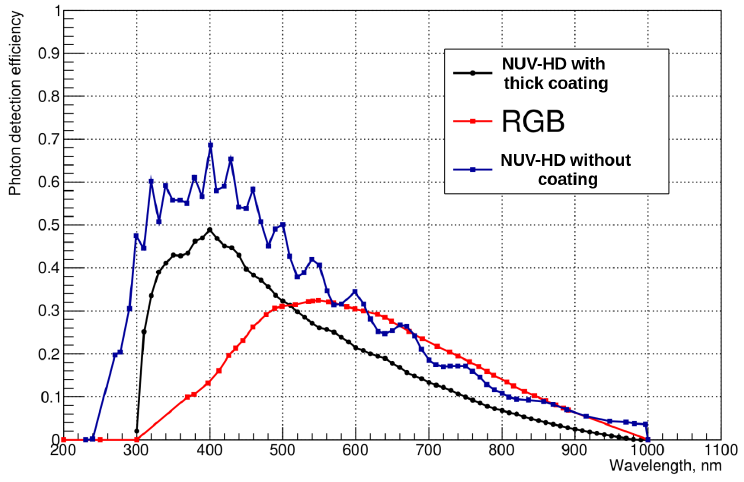}
\includegraphics[scale=.41]{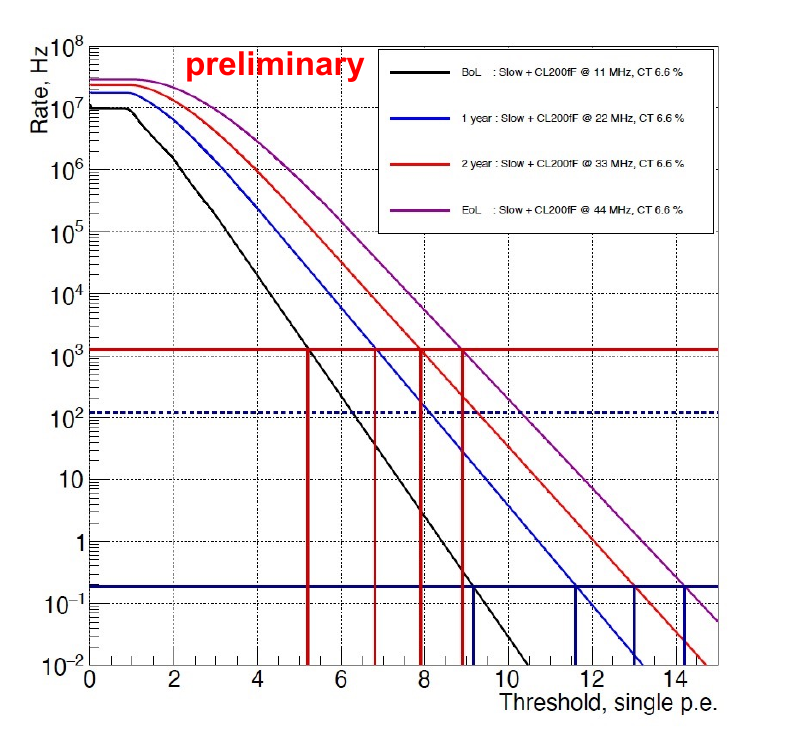}
\caption{[Left panel] Photon detection efficiency versus photon wavelength for different SiPM types by FBK (as labelled). [Right panel] Single pixel trigger rate as a function of the threshold expressed in photo electrons (p.e.) for DCR and NGB values estimated at different times during the mission life (as labelled). The horizontal blue dashed line corresponds to the maximum event rate of 120~Hz, see text. Horizontal blue line corresponds to the maximum single pixel trigger rate (120~Hz/640 $\sim$ 0.18~Hz per pixel). The horizontal red line corresponds to 1.25~kHz (maximum single pixel rate with two pixels cluster in the hit-map). The vertical lines shows thresholds for single (blue) and double coincidence (red) trigger logic.}
\label{fig4}  
\end{figure}
$S = 0.1~\mathrm{m^2}$ is the collecting area of the telescope's primary mirror; $PDE_{eff} = 0.1$ is the total Photon Detection Efficiency (PDE) calculated from the convolution of the SiPM PDE and the NGB spectrum (see figure \ref{fig4} left panel); $\Delta \Omega \simeq (FoV_{pix})^2$ is the pixel viewing solid angle; $\phi_{NGB} = 1.55 \times 10^4 ~m^{-2}sr^{-1}ns^{-1}$ is the total integrated NGB flux in the wavelengths range $\lambda=300~nm$ $\lambda=1000~nm$ \cite{NGB}; $\eta = 6$ is a conservative safety factor that takes into account the expected large fluctuations of the NGB flux. The rate per pixel due to the NGB estimated by these reference values is $R_{pix}^{NGB}\sim$10~MHz. 

The background radiation expected at the operating altitudes of Terzina were estimated using the SPENVIS computation scheme ({\url www.spenvis.oma.be}), focusing on the dominant component of electrons and protons in the 100~keV -  100~MeV energy range, coupled with a Geant4 simulation  ({\url geant4.web.cern.ch/geant4}) of the full detector (mechanical structure, optics and FPA). The effect of the in-orbit background radiation on the FPA is twofold: from one side it can mimic events, as for Cherenkov emission produced by electrons inside the telescope's optical/mechanical parts and hitting on the SiPM layer, on the other side it produces a progressive sensor damage with an increasing Dark Current Rate (DCR) during the mission. 

The in-orbit time evolution of the SiPMs characteristics and their related power consumption are crucial factors that should be taken into account in choosing the sensors technology \cite{UHECR22}. The SiPMs chosen are the NUV-HD series produced by FBK and designed for the near-UV visible wavelengths \cite{SiPM_FBK}. The NUV-HD SiPM technology has typical operating parameters for 35 $\mu$m cell-size given by: DCR $\simeq$ 100 kHz/mm$^2$, after-pulsing AP $\simeq 5\%$ and optical crosstalk CT $\simeq 5\% - 20\%$. In figure \ref{fig4} left panel, we plot the PDE of different SiPM produced by FBK, our baseline solution is the NUV-HD without coating (blue line in left panel of figure \ref{fig4}) \cite{UHECR22,SiPM}. Given the SiPM choice we can simulate the background rate due to NGB and DCR at different times of the Terzina mission: at BoL, after the first and second years and at EoL, the rates obtained are: 11 MHz, 22 MHz, 33 MHz, 44 MHz respectively \cite{UHECR22}. In the right panel of figure \ref{fig4}  we plot the trigger rate per pixel as function of the number of photo-electrons (p.e.) produced by the SiPM at BoL, after one year, after two years and at EoL. It is evident the effect on the expected trigger rate of the increased DCR due to radiation effects.

At EoL, the power consumption of the sensors of the camera, operated at an over-voltage of about 6~V, will reach 0.2~W. This figure does not include the power consumption of the 10 ASICs, which are expected to consume 5~mW/channel \cite{TerzinaProc}, 3.2~W for the 640 channels of the camera. The overall power needed to operate the FPA is expected to be lower than 3.5~W.

The ASIC technology is discussed in \cite{TerzinaProc}, here we recall that each ASIC has 64 channels (8x8 pixels of a single SiPM array), each channel has a memory with a total number of 256 cells (12bit resolution, sampling frequency 200 MHz) arranged in 8 blocks of 32 cells each. Each ASIC has two programmable thresholds (low $S_0$ and high $S_1$) and a clock cycle $T_{{\rm clk}}=5$~ns. The trigger scheme is based on the recognition of specific pixels topologies in the hit-map of the ASIC depending on the (low or high) threshold exceeded by multiple pixels (see \cite{TerzinaProc} for a detailed discussion). If one channel at the time $t_S$ exceeds $S_0$ ($S_1$), during the time interval $\Delta t_c = 16 ~T_{{\rm clk}} = 80$~ns changes to the pixels state are accepted, after $\Delta t_c$ the hit-map is transferred to the FPGA. The event is accepted by the FPGA if the hit-map shows two (three) or more adjacent pixels, with defined topologies, above $S_1$ ($S_0$).  Once the event is accepted it will be centred for digitisation at $t_S$ and digitised through 32 time-samples of the signal spaced by $T_{{\rm clk}}$ for a total sampled time interval $2\Delta t_c=160$~ns in the time interval ($t_S-\Delta t_c, t_S+\Delta t_c$), occupying one memory block per channel (pixel). The hit-map recognition chain has a total duration of 250~ns, given by $\Delta t_c=80$~ns plus the time needed to transfer the hit-map to the FPGA (140~ns) and the time needed by the FPGA to recognise the event and communicating it back to the ASIC (30~ns). Each ASIC is able to manage in parallel a maximum of 8 (number of memory blocks per channel) digitisation processes \cite{TerzinaProc}.  

The digitised signal in a single pixel is encoded in a number of bit: 12x32+header+padding = 434 bit. Thus, an event in a single ASIC is encoded in 64x434 = 27776 bit. The maximum downlink data stream for Terzina is 45~Gbit/day, which corresponds to an absolute maximum number of events per day that can be sent for the offline analysis $1.29\times 10^{7}$ events/day, roughly an event rate of 150 Hz. %In the following we will assume an actual event rate of 120 Hz, taking into account the downlink needs of telemetry and platform control.  
The possibility of reprogramming the ASICs thresholds during the flight guarantees the opportunity to adjust $S_0$ and $S_1$ to the changing response of the sensors, in order to maintain a fixed event rate. In the right panel of figure \ref{fig4}, we show the variation of $S_0$ (red vertical lines) and $S_1$ (blu vertical lines) with the increasing mission age. These estimations follows by defining $S_0$ as the threshold corresponding to the maximum single pixel rate $120/640$~Hz (blu horizontal line) and $S_1$ as the threshold corresponding to the maximum allowed single pixel rate in the case of configurations with two pixels clustering 1.25 kHz (red horizontal line) as discussed in \cite{UHECR22}. 

\section{Conclusions} 
\label{sec:Conclusions}

In conclusion, we provide a preliminary estimation of the detection capabilities of the Terzina instrument. When monitoring below the limb, Terzina will perform background sampling with hit-map recording at a rate that can reach several Hz. From above the limb, Terzina is capable of observing CR events. 
\begin{figure}[!h]
\centering
\includegraphics[scale=.38]{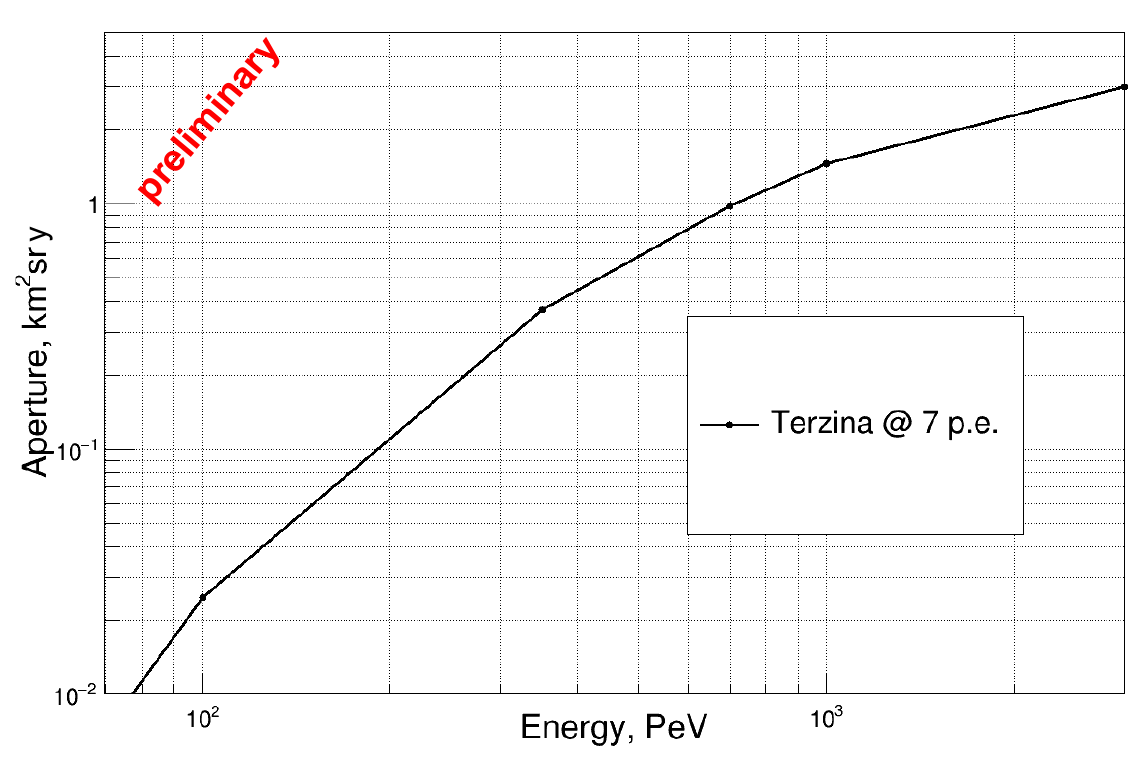}
\caption{Expected Terzina aperture to protons' EAS vs energy during the first year of operation.}
\label{fig5} 
\end{figure}
Figure 5 illustrates the expected detector's aperture associated with CR protons, calculated based on the BoL sensors' response and a single threshold trigger scheme with $S_0=7$ p.e.  The results in figure \ref{fig5} are obtained by combining the Geant4 simulation scheme of the detector with the EASCherSim computation scheme for generating protons' EAS. Considering the observed CR flux at energies exceeding 100 PeV, $\phi_{CR}\sim 6.6\times 10^3$ ${\rm km}^{-2}{\rm sr}^{-1}{\rm y}^{-1}$ \cite{Auger}, assuming a proton fraction of 50$\%$ \cite{Auger}, and a detector duty cycle of $40\%$, the aperture shown in figure \ref{fig5} demonstrates Terzina's capability to observe a significant number of CR proton events (with $E\ge100$ PeV) already during the first year of operation, with an estimated count of no less than 20 events per year. Achieving this level of detection would serve as a clear validation of the experimental technique employed by Terzina.

\vskip 0.5 cm 

\noindent {\bf{Acknowledgements}}. 
%\section*{Acknowledgements}
NUSES is funded by the Italian Government (CIPE n. 20/2019), by the Italian Minister of Economic Development (MISE reg. CC n. 769/2020), by the Italian Space Agency (CDA ASI n. 15/2022), by the Swiss National Foundation (SNF grant n. 178918) and by the European Union - NextGenerationEU under the MUR National Innovation Ecosystem grant ECS00000041 - VITALITY - CUP D13C21000430001.

\clearpage

\section*{Full Authors List NUSES Collaboration}
\small

\begin{sloppypar}\noindent
%\documentclass[onecolumn]{article}

%\begin{document}
%\title{{\huge The NUSES Collaboration}}
%\maketitle
%\begin{center}
\noindent
R. Aloisio$^{1,2}$,
C. Altomare$^{3}$,
F. C. T. Barbato$^{1,2}$,
R. Battiston$^{4,5}$,
M. Bertaina$^{6,7}$,
E. Bissaldi$^{3,8}$,
D. Boncioli$^{2,9}$,
L. Burmistrov$^{10}$,
I. Cagnoli$^{1,2}$,
M. Casolino$^{11,12}$,
A.L. Cummings$^{13}$,
N. D'Ambrosio$^{2}$,
I. De Mitri$^{1,2}$,
G. De Robertis$^{3}$,
C. De Santis$^{11}$,
A. Di Giovanni$^{1,2}$,
A. Di Salvo$^{7}$,
M. Di Santo$^{1,2}$,
L. Di Venere$^{3}$,
J. Eser$^{14}$,
M. Fernandez Alonso$^{1,2}$,
G. Fontanella$^{1,2}$,
P. Fusco$^{3,8}$,
S. Garbolino$^{7}$,
F. Gargano$^{3}$,
R. A. Giampaolo$^{1,7}$,
M. Giliberti$^{3,8}$,
F. Guarino$^{15,16}$,
M. Heller$^{10}$,
R. Iuppa$^{4,5}$,
J. F. Krizmanic$^{17,18}$,
A. Lega$^{4,5}$,
F. Licciulli$^{3}$,
F. Loparco$^{3,8}$,
L. Lorusso$^{3,8}$,
M. Mariotti$^{19,20}$,
M. N. Mazziotta$^{3}$,
M. Mese$^{15,16}$,
H. Miyamoto$^{1,7}$,
T. Montaruli$^{10}$,
A. Nagai$^{10}$,
R. Nicolaidis$^{4,5}$,
F. Nozzoli$^{4,5}$,
A. V. Olinto$^{14}$,
D. Orlandi$^{2}$,
G. Osteria$^{15}$,
P. A. Palmieri$^{6,7}$,
B. Panico$^{15,16}$,
G. Panzarini$^{3,8}$,
A. Parenti$^{1,2}$,
L. Perrone$^{21,22}$,
P. Picozza$^{12,11}$,
R. Pillera$^{3,8}$,
R. Rando$^{19,20}$,
M. Rinaldi$^{11}$,
A. Rivetti$^{7}$,
V. Rizi$^{2,9}$,
F. Salamida$^{2,9}$,
E. Santero Mormile$^{6}$,
V. Scherini$^{21,22}$,
V. Scotti$^{15,16}$,
D. Serini$^{3}$,
I. Siddique$^{1,2}$,
L. Silveri$^{1,2}$,
A. Smirnov$^{1,2}$,
R. Sparvoli$^{11}$,
S. Tedesco$^{7,23}$,
C. Trimarelli$^{10}$,
L. Wu$^{1,2,\dag}$,
P. Zuccon$^{4,5}$,
S. C. Zugravel$^{7,23}$.
\small{
\vskip 1.0cm\noindent
$^1$Gran Sasso Science Institute (GSSI), Via Iacobucci 2, I-67100 L'Aquila,  Italy\\
$^2$Istituto Nazionale di Fisica Nucleare (INFN) - Laboratori Nazionali del Gran Sasso, I-67100 Assergi, L'Aquila, Italy\\
$^{3}$Istituto Nazionale di Fisica Nucleare, Sezione di Bari, via Orabona 4, I-70126 Bari, Italy\\
$^4$Dipartimento di Fisica, Università di Trento, via Sommarive 14 I-38123 Trento, Italy\\
$^5$Istituto Nazionale di Fisica Nucleare (INFN) - TIFPA, via Sommarive 14 I-38123 Trento, Italy\\
$^6$Dipartimento di Fisica, Università di Torino, Via P. Giuria, 1 I-10125 Torino, Italy\\
$^7$Istituto Nazionale di Fisica Nucleare (INFN) - Sezione di Torino, I-10125 Torino, Italy\\
$^{8}$Dipartimento di Fisica M. Merlin, dell’Università e del Politecnico di Bari, via Amendola 173, I-70126 Bari, Italy\\
$^9$Dipartimento di Scienze Fisiche e Chimiche, Università degli Studi di L'Aquila, I-67100 L'Aquila, Italy\\
$^{10}$Département de Physique Nuclèaire et Corpusculaire, Université de Genève, 1205 Genève, Switzerland\\
$^{11}$INFN Roma Tor Vergata, Dipartimento di Fisica, Universitá di Roma Tor Vergata, Roma, Italy\\
$^{12}$RIKEN, 2-1 Hirosawa, Wako, Saitama, Japan\\
$^{13}$Departments of Physics and Astronomy \& Astrophysics, Institute for Gravitation and the Cosmos, Pennsylvania State University, University Park, PA 16802, USA\\
$^{14}$Department of Astrophysics \& Astronomy, The University of Chicago, Chicago, IL 60637, USA\\
$^{15}$Istituto Nazionale di Fisica Nucleare, Sezione di Napoli, via Cintia, I-80126 Napoli, Italy\\
$^{16}$Dipartimento di Fisica E. Pancini dell'Università di Napoli Federico II, via Cintia, I-80126 Napoli, Italy\\
$^{17}$CRESST/NASA Goddard Space Flight Center, Greenbelt, MD 20771, USA\\
$^{18}$University of Maryland, Baltimore County, Baltimore, MD 21250, USA\\
$^{19}$Università di Padova, I-35122 Padova, Italy\\
$^{20}$Istituto Nazionale di Fisica Nucleare (INFN) - Sezione di Padova, I-35131 Padova, Italy\\
$^{21}$Dipartimento di Matematica e Fisica ``E. De Giorgi", Università del Salento, Via per Arnesano, I-73100 Lecce, Italy\\
$^{22}$Istituto Nazionale di Fisica Nucleare - INFN - Sezione di Lecce, Via per Arnesano, I-73100 Lecce, Italy\\
$^{23}$Department of Electrical, Electronics and Communications Engineering, Politecnico di Torino, Corso Duca degli Abruzzi 24, I-10129 Torino, Italy\\
$^\dag$Now at Institute of Deep Space Sciences, Deep Space Exploration Laboratory, Hefei 230026, China
}
\end{sloppypar}

%% Full authors list (ONLY FOR COLLABORATIONS)
%\clearpage
%\section*{Full Authors List: \Coll\ The NUSES Collaboration}
%
%\noindent \textbf{Note comment afterwards:} Collaborations have the possibility to provide an authors list in xml format which will be used while generating the DOI entries making the full authors list searchable in databases like Inspire HEP. \\
%
%\scriptsize
%\noindent
%first.author$^1$, 
%second.author$^2$, 
%third.author$^3$ % .... more names
%and 
%last.author$^{n}$ \\
%
%\noindent
%$^1$first.affiliation.
%$^2$second.affiliation. % .... more affiliation
%$^{m}$last.affiliation.

\end{document}